\documentclass[twocolumn]{aastex631}

\shorttitle{Chance coincidences of LMXBs and SNRs}
\shortauthors{Heinke}

\begin{document}

\title{Chance coincidences between black hole low-mass X-ray binaries and supernova remnants}



\author[0000-0003-3944-6109]{Craig O. Heinke}
\affiliation{Physics Department, CCIS 4-183, University of Alberta, Edmonton, AB, T6G 2E1, Canada}
\email{heinke@ualberta.ca}

\begin{abstract}

I argue that black hole low-mass X-ray binaries (BH LMXBs) are very unlikely to be physically associated with supernova remnants (SNRs). The timescales of BH LMXBs are so much longer than those of SNRs, that there is only a 0.2\% chance of any BH LMXB being identified within its natal SNR. However, the probability of a BH LMXB being projected within a SNR is significant; I estimate that 2 BH LMXBs should be projected within SNRs from our perspective.  I look more closely at the suggestion by Balakrishnan and collaborators of an association between the BH  X-ray binary Swift J1728.9-3613 and the SNR G351.9-0.9, and show that this is most likely a chance coincidence.

\end{abstract}



\section{Introduction} \label{sec:intro}

Identification of compact objects with their natal supernova remnants is crucial for understanding both the supernova process and the properties of the remmants.  A key question is whether the formation of black holes generally produces a standard supernova remnant (SNR), or instead whether many supernovae producing black holes 'fail' \citep[e.g.][]{Kochanek08}. Identification of black holes in SNRs has been harder than neutron stars, because young neutron stars radiate while black holes do not; suggestions of young solitary black holes in SNRs depend on a lack of candidate neutron stars \citep{Kaplan04,Kaplan06,Lopez13}. 

Binary black holes may exist in SNRs, and evidence has accumulated (e.g. the morphology of the SNR shell appears to be directly affected by the observed jets) that SS 433 is a high-mass  black hole X-ray binary located in a SNR  \citep{ClarkMurdin78,Margon84,Bowler18,Cherepashchuk20}. However, there are relatively few black holes known in high-mass X-ray binaries, while there are a number of black holes in low-mass X-ray binaries (LMXBs). Identifying black hole LMXBs with SNRs would be very useful for understanding the supernova process. Recently, there have been a few claims of association of black hole LMXBs with SNRs \citep{Maxted2020,Balakrishnan23}, which motivate  discussion.

SNRs occupy a significant fraction of portions of the Galactic Plane, so there is a long history of questioning whether projections are really associated (e.g. \citealt{Gaensler01} critically discussed claims of SNR associations with magnetars). In this article, I consider two probabilities; a) the probability that any black hole LMXBs in our Galaxy would be observed while still inside their natal SNRs; and b) the probability that black hole LMXBs would be projected by chance onto unrelated SNRs. I then consider the specific case of Swift J1728.9-3613
 and G351.9-0.9, advanced by \citet{Balakrishnan23}.

\section{Evolutionary considerations}

How likely is it that a BH LMXB should be found within its native SNR, a priori? We can make a very straightforward estimate of this probability using their respective lifetimes.  
Typical SNRs in the Galaxy are observable in the radio for up to about 30,000 years \citep{Leahy2020}.
Black hole LMXBs are believed to be mass-transferring for of order 1 Gyr (e.g. Fig. 9 of Podsiadlowski et al. 2003; though this is clear from the typical mass transfer rates of $\sim10^{-9}$ Msun/year).  
It is possible for this timescale to be up to a factor of 10 shorter for a portion of BH LMXBs if irradiation-induced mass-transfer cycles are active (see \citealt{Pfahl03,Buning04}). 
This does not include the time between the BH formation and the beginning of mass transfer, which can be several Gyrs; let us take 1 Gyr as a typical lower limit. Thus, an average BH LMXB has an age of $>$0.5 Gyr.  

If we know of roughly 75 BH LMXBs in the Galaxy (as catalogued by e.g. \citealt{Corral-Santana16,Tetarenko16}), then typically the youngest one will have an age of $\sim$10-15 million years  (assuming the present time is not special). The probability of seeing any of 75 BH LMXBs in our Galaxy still in its natal SNR should then have a probability of $75\times(30,000/10^9$) $\sim$0.002 or 0.2\%. This is an upper limit, due to the likelihood that most BH LMXBs spend some Gyrs before beginning mass transfer. Thus, it is highly unlikely that we will see any bright BH LMXB within its natal SNR.

Note that the probability is similarly low for a neutron star LMXB to be observed within a SNR, as they have similar numbers and mass transfer rates. On the other hand,  wind-accreting high-mass X-ray binaries have lifetimes of order 1 Myr, rather than 1 Gyr, so the expectation value of the number of high-mass X-ray binaries associated with SNRs, given 169 known high-mass X-ray binaries \citep{Neumann23}, is $169\times(30,000/10^6)$ $\sim$5, consistent with a small number of known high mass X-ray binary/SNR associations in our Galaxy \citep{ClarkMurdin78,Heinz13}, and in the LMC  \citep{Maitra19,Maitra21}. \citet{Xing21} perform a more thorough population analysis,  indeed predicting that a few Be X-ray binaries should be associated with SNRs. 

\section{Chance coincidence on the sky}

I performed a straightforward calculation for an order of magnitude estimate of a chance coincidence. I utilized the fact that SNRs and BH LMXBs are both rather concentrated towards the central part of our Galaxy. This is likely for different reasons; BH LMXBs are concentrated in the bulge (as an old population), while SNRs are concentrated in the inner star-forming ring (as a young population), but both show a  concentration towards the Galactic Plane in the direction of the Galactic Bulge. 

To keep things simple, I used the band $\pm$2 degrees in Galactic Latitude. There are 234 SNRs within 2 degrees of the Plane (according to the Green catalogue, \citet{Green19} (\url{http://www.mrao.cam.ac.uk/surveys/snrs/snrs.data.html}). 
I restrict to $\pm$50 degrees in Galactic longitude around the Galactic Center as well; this gives 200 square degrees in which there are 168 SNRs (from the Green catalogue) and 22 transient BH LMXBs (according to the BlackCat catalogue, \url{https://www.astro.puc.cl/BlackCAT/transients.php}).  Other selections of Galactic longitude and latitude are possible, but increasing the limits substantially would decrease the number of predicted chance coincidences; this selection encompasses an area where both BH LMXBs and SNRs are concentrated on the sky.

   I selected the 168 SNRs in this part of the sky from the Green catalog, and estimated their areas as $\pi*(D/2)^2$ from the D values given (where the SNRs are elliptical, I chose a median, rounding down to the nearest arcminute). This gives a total of 18 square degrees covered by SNRs within this 200 square degree region, so a 9\% chance of any given BH LMXB being projected inside a SNR. Given that there are 22 BH LMXBs in this region, we might expect of order 2 BH LMXBs to be projected within SNRs in this portion of the sky. And indeed two matches have been reported; G323.7-1.0/MAXI J1535-571 by \citet{Maxted2020}, and G351.9-0.9/Swift J1728.9-3613 by \citet{Balakrishnan23}. I argue that both of these are chance coincidences. I will focus on the latter  work here, as it has been peer-reviewed. 

\section{Is Swift J1728.9-3613 associated with G351.9-0.9?}

Recently, \citet{Balakrishnan23} argued that Swift J1728.9-3613 is associated with G351.9-0.9, using a suite of galaxy simulations, with simulated SNRs and BH LMXBs, to argue that the chance of projection, considering the distance estimates they make for both objects, was very low. I argue that they have underestimated the probability of a chance coincidence, and have overestimated the impact of the distance constraints they use, and that the probability of an actual connection is very low. 


\citet{Balakrishnan23} make the following key arguments. 
Swift J1728.9-3613 is a BH  X-ray binary (based principally on its X-ray colors).
Swift J1728.9-3613 is located at 8.4$\pm0.8$ kpc (using spectral maps of atomic and molecular gas to infer distance from column density). 
They identify an infrared counterpart, based on brightening at the time of the X-ray outburst,  which they identify as an intermediate-mass donor star.
They argue from the X-ray non-detection of G359.1-0.9 that the SNR's distance exceeds 7.5 kpc (based on a Sedov model of the expected SNR brightness).
\citet{Balakrishnan23} then perform 10,000 simulations of Milky Way-like galaxies, using  distributions of SNRs and BH LMXBs derived from the \citet{Green19} and \citet{Corral-Santana16} catalogues, to estimate a) the probability of a BH  X-ray binary being coincidentally projected within a SNR; and b) the probability of such a projection, where the distances of the SNR and BH X-ray binary match the ranges identified for G351.9-0.9 and Swift J1728.9-3613, respectively. 
For a), they only consider overlaps if the BH  X-ray binary is projected within $r/2$ of the SNR's centre, and find an average of 0.52 overlaps per galaxy. 
For b), the added requirements of satisfying the distance constraints give them an average of 0.007 overlaps per galaxy, and they infer from this a probability of chance superposition of $<$1.7\% (at 99.7\% confidence). 

\begin{figure}[ht!]
	\includegraphics[width=3.5in]{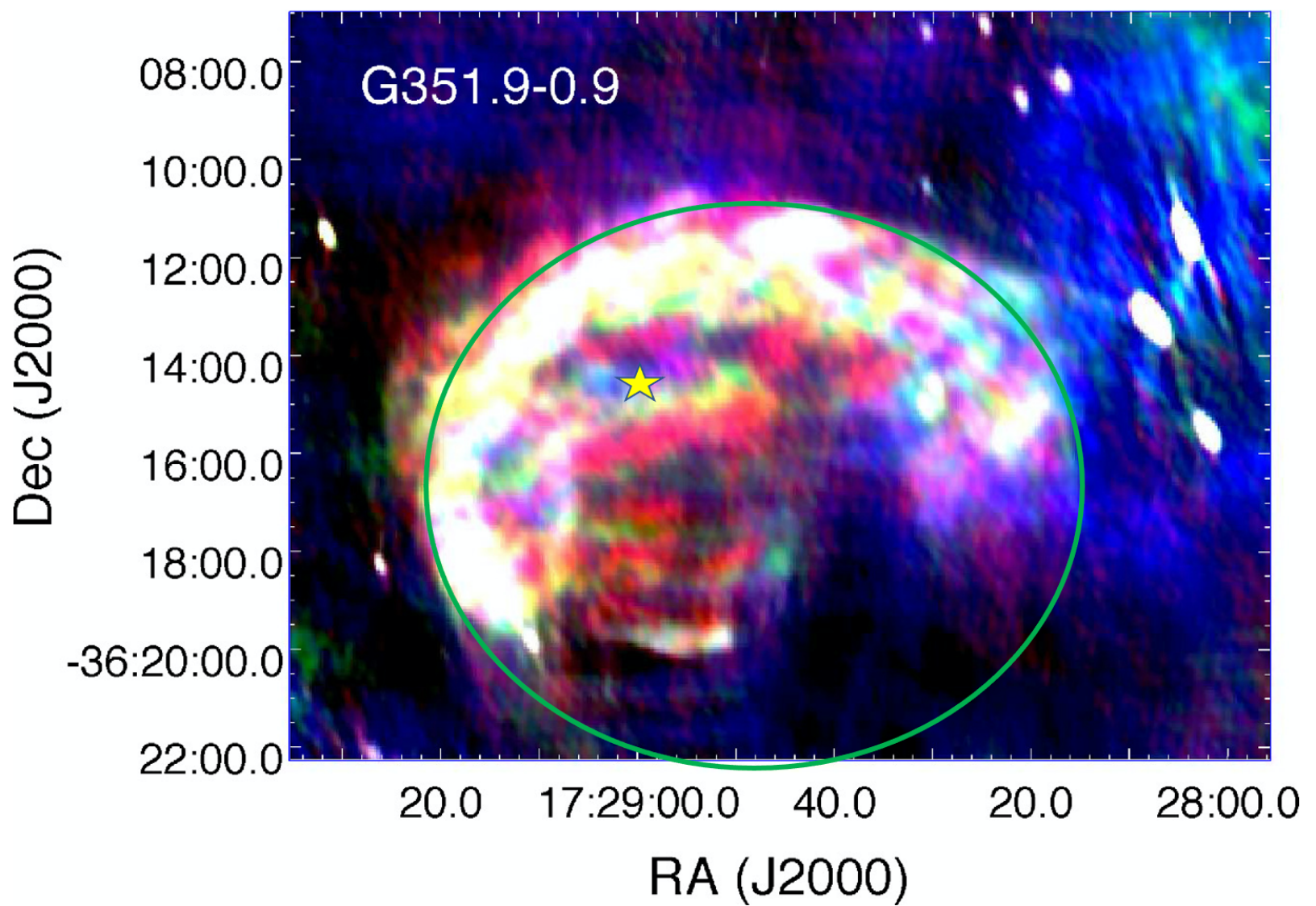}
    \caption{Combined radio image of G351.9-0.9 in 321 MHz (red), 385 MHz (green), and 450 MHz (blue), from \citet{Veena19}. The yellow star marks the location of Swift J1728.9-3613. This image emphasizes the partial shell nature of G351.9-0.9, and suggests the SNR extends, unseen, to the south.  The green ellipse suggests a possible symmetric expansion of the SNR. If so, Swift J1728.9-3613 seems not to be located within half the SNR's radius of its center.}
    \label{fig:Veena}
\end{figure}

One concern is the assumption by Balakrishnan et al. that Swift J1728.9-3613 is located within half the radius of G351.9-0.9, which seems to be based on the assumption that the full extent of G351.9-0.9 is visible in their Fig. 5. This is less obvious to me, as other radio images of G351.9-0.9 suggest that we see only part of the shell of this SNR (see Fig.\ref{fig:Veena})  Partial SNR shells are common (see e.g. G348.5+0.1, G348.5-0.0, G348.7+0.3, G338.1+0.4, G315.9-0.0, G304.6+0.1, G338.1+0.4, G299.6-0.5, G317.3-0.2, and 
G327.4+1.0, as well as G351.9-0.9, in \citealt{Whiteoak96}); if we are indeed seeing a partial SNR shell, then Swift J1728.9-3613 may not be located within half the radius of the center of the explosion. 

\citet{Balakrishnan23} argue that their infrared counterpart in quiescence is consistent with an A or B-type star, although they note that the colors do not match such a star perfectly. However, their calculation of the interstellar extinction does not match the {\it wilm} \citep{Wilms01} abundance tables they use to measure $N_H$ from X-ray spectroscopy. For the {\it wilm} abundances, the empirical conversion from $A_V$ to $N_H$ is $2.81--2.87\times10^{21}$ $A_V$ cm$^{-2}$ \citep{Bahramian15,Foight16}, which gives an $A_V$ of 14, rather than the $A_V$=22 they derived from $N_H=4\times10^{22}$. This is important, because fixing this error makes the inferred companion star much less luminous at the same assumed distance.

Using $A_K/A_V$=0.108 \citep[for $R_V=3.1$,][]{Cox00}, I find $A_K$=1.5. \citet{Balakrishnan23} measure $K=18.2$ in quiescence, so $K_0=16.7$, and for an assumed distance $d=8.4$ (7.6-9.2) kpc, $M_K=2.1$ (1.9-2.3). This is consistent with an F0 main sequence star, for which $M_V=2.7$ and $V-K=0.7$ \citep{Cox00}. Considering the $J-K$ color, I calculate an infrared reddening $E(J-K)=2.4$, and thus an intrinsic $J-K=0.3$, consistent with an F5 star. B9 stars have $J-K=0$; reaching this color would require increasing $N_H$ to $4.5\times10^{22}$ (perhaps not implausible), while to reach a B9 absolute magnitude of $\sim$0 would require that the star be at a distance of $\sim$20 kpc. At 20 kpc, and a Galactic latitude of -0.96 degrees, this would require the B9 star to be 335 pc below the plane. As the scale height of A stars is only 50-90 pc \citep{Bovy17,Xiang18}, and the Milky Way warps toward +Z, not -Z, at this location \citep{Skowron19}, it is unlikely that a young B star would be located there. 
This argument shows that the companion is a low-mass star, and thus that the system is definitely an LMXB. (This result is important, because a high-mass X-ray binary would have a shorter lifetime, affecting the evolutionary argument above.)


I agree with \citet{Balakrishnan23} that, if we assume that a BH-LMXB/SNR overlap must require the BH LMXB to be located within half the SNR radius of the center, the average number of predicted overlaps in the Milky Way should be approximately 0.5. (My estimate of 2 overlaps above does not require the BH LMXB to be located within half the SNR radius.) 

 When computing their final chance coincidence estimate,  \citet{Balakrishnan23} also consider distance information, ending with a probability of chance superposition of $<$1.7\% at 99.7\% confidence.  
However, they did not filter to see whether a simulated SNR and BH LMXB  have consistent distance estimates, but rather for objects with the specific distances inferred for these objects ($>$7.5 kpc for G351.9-0.9, 8.4$\pm$0.8 kpc for Swift J1728.9-3613). This seems to be an incorrect choice; 
I think 
that this pairing would have been suggested as a match regardless of the specific distances, if the distances were compatible. 

Assuming the distance constraints that \citet{Balakrishnan23} infer, I can ask, what is the probability that a BH LMXB happening to fall along the line of sight towards this SNR would fall within the distance range given for G351.9-0.9 ($>$7.5 kpc)?   
I can answer this question by calculating the fraction of BH LMXBs that have estimated distances $>$7.5 kpc, lying within 
90 degrees in longitude of the Galactic Centre, with no latitude limit (the same range considered by \citealt{Balakrishnan23}).  The BlackCAT catalogue lists 36
 BH transients with distance estimates in the Balakrishnan range, of which 22 
are consistent with a distance $>$7.5 kpc. 
Within 60 degrees in longitude and 2 degrees in latitude (the range considered in \S 3), I find 11 BH transients with distance estimates, of which 7 are consistent with $d>$7.5 kpc, so this is a similar fraction.
Thus, I estimate that applying the distance constraint reduces the probability of a chance coincidence by a factor of 22/36. If a BH LMXB must be within half the SNR radius of the centre of a SNR to count as an overlap, the expectation value of overlaps then becomes 0.31; if the half-radius requirement is not applied, the expectation value of overlaps becomes 1.2.

An inferred chance coincidence match rate of 0.31, or 1.2, per Milky Way, including the effect of the distance constraint, is dramatically larger than the value of $<$1.7\% at 99.7\% confidence stated by \citet{Balakrishnan23}. Combined with my a priori expectation that identification of a BH LMXB with a SNR is highly unlikely (\S 2), I conclude that this pairing is almost certainly a chance coincidence.

\section{Conclusions}

Based on the lifetimes of SNRs and black hole LMXBs, the probability of finding one true association between a black hole LMXB and a SNR in the Milky Way is about 0.2\%. On the other hand, the expected number of black hole LMXBs projected within SNRs, in the crowded Galactic Plane towards the Bulge, is of order 2--that is, a chance coincidence is $\sim$1000 times more likely.  \citet{Balakrishnan23} argue that the black hole  X-ray binary Swift J1728.9-3613 is associated with the SNR G351.9-0.9, but their weak distance constraint on SNR G351.9-0.9 does not substantially reduce the probability of a chance coincidence, so I argue that this match is spurious. 

\begin{acknowledgments}

I thank Sharon Morsink, Tom Maccarone,  \& Greg Sivakoff for helpful conversations, and Veena Vadamattom for permission to reproduce Fig. 1. I am supported by NSERC Discovery Grant RGPIN 2016-04602.

\end{acknowledgments}


\bibliography{sample631}{}

\begin{thebibliography}{}
\expandafter\ifx\csname natexlab\endcsname\relax\def\natexlab#1{#1}\fi
\providecommand{\url}[1]{\href{#1}{#1}}
\providecommand{\dodoi}[1]{doi:~\href{http://doi.org/#1}{\nolinkurl{#1}}}
\providecommand{\doeprint}[1]{\href{http://ascl.net/#1}{\nolinkurl{http://ascl.net/#1}}}
\providecommand{\doarXiv}[1]{\href{https://arxiv.org/abs/#1}{\nolinkurl{https://arxiv.org/abs/#1}}}

\bibitem[{{Bahramian} {et~al.}(2015){Bahramian}, {Heinke}, {Degenaar},
  {Chomiuk}, {Wijnands}, {Strader}, {Ho}, \& {Pooley}}]{Bahramian15}
{Bahramian}, A., {Heinke}, C.~O., {Degenaar}, N., {et~al.} 2015, \mnras, 452,
  3475, \dodoi{10.1093/mnras/stv1585}

\bibitem[{{Balakrishnan} {et~al.}(2023){Balakrishnan}, {Draghis}, {Miller},
  {Bright}, {Fender}, {Ng}, {Cackett}, {Fabian}, {Kuntz}, {Miller-Jones},
  {Proga}, {Ray}, {Raymond}, {Reynolds}, \& {Zoghbi}}]{Balakrishnan23}
{Balakrishnan}, M., {Draghis}, P.~A., {Miller}, J.~M., {et~al.} 2023, arXiv
  e-prints, arXiv:2303.04159, \dodoi{10.48550/arXiv.2303.04159}

\bibitem[{{Bovy}(2017)}]{Bovy17}
{Bovy}, J. 2017, \mnras, 470, 1360, \dodoi{10.1093/mnras/stx1277}

\bibitem[{{Bowler} \& {Keppens}(2018)}]{Bowler18}
{Bowler}, M.~G., \& {Keppens}, R. 2018, \aap, 617, A29,
  \dodoi{10.1051/0004-6361/201732488}

\bibitem[{{B{\"u}ning} \& {Ritter}(2004)}]{Buning04}
{B{\"u}ning}, A., \& {Ritter}, H. 2004, \aap, 423, 281,
  \dodoi{10.1051/0004-6361:20035678}

\bibitem[{{Cherepashchuk} {et~al.}(2020){Cherepashchuk}, {Postnov}, {Molkov},
  {Antokhina}, \& {Belinski}}]{Cherepashchuk20}
{Cherepashchuk}, A., {Postnov}, K., {Molkov}, S., {Antokhina}, E., \&
  {Belinski}, A. 2020, \nar, 89, 101542, \dodoi{10.1016/j.newar.2020.101542}

\bibitem[{{Clark} \& {Murdin}(1978)}]{ClarkMurdin78}
{Clark}, D.~H., \& {Murdin}, P. 1978, \nat, 276, 44, \dodoi{10.1038/276044a0}

\bibitem[{{Corral-Santana} {et~al.}(2016){Corral-Santana}, {Casares},
  {Mu{\~n}oz-Darias}, {Bauer}, {Mart{\'\i}nez-Pais}, \&
  {Russell}}]{Corral-Santana16}
{Corral-Santana}, J.~M., {Casares}, J., {Mu{\~n}oz-Darias}, T., {et~al.} 2016,
  \aap, 587, A61, \dodoi{10.1051/0004-6361/201527130}

\bibitem[{{Cox}(2000)}]{Cox00}
{Cox}, A.~N. 2000, {Allen's astrophysical quantities}

\bibitem[{{Foight} {et~al.}(2016){Foight}, {G{\"u}ver}, {{\"O}zel}, \&
  {Slane}}]{Foight16}
{Foight}, D.~R., {G{\"u}ver}, T., {{\"O}zel}, F., \& {Slane}, P.~O. 2016, \apj,
  826, 66, \dodoi{10.3847/0004-637X/826/1/66}

\bibitem[{{Gaensler} {et~al.}(2001){Gaensler}, {Slane}, {Gotthelf}, \&
  {Vasisht}}]{Gaensler01}
{Gaensler}, B.~M., {Slane}, P.~O., {Gotthelf}, E.~V., \& {Vasisht}, G. 2001,
  \apj, 559, 963, \dodoi{10.1086/322358}

\bibitem[{{Green}(2019)}]{Green19}
{Green}, D.~A. 2019, VizieR Online Data Catalog, VII/284

\bibitem[{{Heinz} {et~al.}(2013){Heinz}, {Sell}, {Fender}, {Jonker}, {Brandt},
  {Calvelo-Santos}, {Tzioumis}, {Nowak}, {Schulz}, {Wijnands}, \& {van der
  Klis}}]{Heinz13}
{Heinz}, S., {Sell}, P., {Fender}, R.~P., {et~al.} 2013, \apj, 779, 171,
  \dodoi{10.1088/0004-637X/779/2/171}

\bibitem[{{Kaplan} {et~al.}(2004){Kaplan}, {Frail}, {Gaensler}, {Gotthelf},
  {Kulkarni}, {Slane}, \& {Nechita}}]{Kaplan04}
{Kaplan}, D.~L., {Frail}, D.~A., {Gaensler}, B.~M., {et~al.} 2004, \apjs, 153,
  269, \dodoi{10.1086/421065}

\bibitem[{{Kaplan} {et~al.}(2006){Kaplan}, {Gaensler}, {Kulkarni}, \&
  {Slane}}]{Kaplan06}
{Kaplan}, D.~L., {Gaensler}, B.~M., {Kulkarni}, S.~R., \& {Slane}, P.~O. 2006,
  \apjs, 163, 344, \dodoi{10.1086/501441}

\bibitem[{{Kochanek} {et~al.}(2008){Kochanek}, {Beacom}, {Kistler}, {Prieto},
  {Stanek}, {Thompson}, \& {Y{\"u}ksel}}]{Kochanek08}
{Kochanek}, C.~S., {Beacom}, J.~F., {Kistler}, M.~D., {et~al.} 2008, \apj, 684,
  1336, \dodoi{10.1086/590053}

\bibitem[{{Leahy} {et~al.}(2020){Leahy}, {Ranasinghe}, \&
  {Gelowitz}}]{Leahy2020}
{Leahy}, D.~A., {Ranasinghe}, S., \& {Gelowitz}, M. 2020, \apjs, 248, 16,
  \dodoi{10.3847/1538-4365/ab8bd9}

\bibitem[{{Lopez} {et~al.}(2013){Lopez}, {Ramirez-Ruiz}, {Castro}, \&
  {Pearson}}]{Lopez13}
{Lopez}, L.~A., {Ramirez-Ruiz}, E., {Castro}, D., \& {Pearson}, S. 2013, \apj,
  764, 50, \dodoi{10.1088/0004-637X/764/1/50}

\bibitem[{{Maitra} {et~al.}(2021){Maitra}, {Haberl}, {Maggi}, {Kavanagh},
  {Vasilopoulos}, {Sasaki}, {Filipovi{\'c}}, \& {Udalski}}]{Maitra21}
{Maitra}, C., {Haberl}, F., {Maggi}, P., {et~al.} 2021, \mnras, 504, 326,
  \dodoi{10.1093/mnras/stab716}

\bibitem[{{Maitra} {et~al.}(2019){Maitra}, {Haberl}, {Filipovi{\'c}},
  {Udalski}, {Kavanagh}, {Carpano}, {Maggi}, {Sasaki}, {Norris}, {O'Brien},
  {Hotan}, {Lenc}, {Szyma{\'n}ski}, {Soszy{\'n}ski}, {Poleski}, {Ulaczyk},
  {Pietrukowicz}, {Koz{\l}owski}, {Skowron}, {Mr{\'o}z}, {Rybicki}, {Iwanek},
  \& {Wrona}}]{Maitra19}
{Maitra}, C., {Haberl}, F., {Filipovi{\'c}}, M.~D., {et~al.} 2019, \mnras, 490,
  5494, \dodoi{10.1093/mnras/stz2831}

\bibitem[{{Margon}(1984)}]{Margon84}
{Margon}, B. 1984, \araa, 22, 507, \dodoi{10.1146/annurev.aa.22.090184.002451}

\bibitem[{{Maxted} {et~al.}(2020){Maxted}, {Ruiter}, {Belczynski},
  {Seitenzahl}, \& {Crocker}}]{Maxted2020}
{Maxted}, N.~I., {Ruiter}, A.~J., {Belczynski}, K., {Seitenzahl}, I.~R., \&
  {Crocker}, R.~M. 2020, arXiv e-prints, arXiv:2010.15341,
  \dodoi{10.48550/arXiv.2010.15341}

\bibitem[{{Neumann} {et~al.}(2023){Neumann}, {Avakyan}, {Doroshenko}, \&
  {Santangelo}}]{Neumann23}
{Neumann}, M., {Avakyan}, A., {Doroshenko}, V., \& {Santangelo}, A. 2023, arXiv
  e-prints, arXiv:2303.16137, \dodoi{10.48550/arXiv.2303.16137}

\bibitem[{{Pfahl} {et~al.}(2003){Pfahl}, {Rappaport}, \&
  {Podsiadlowski}}]{Pfahl03}
{Pfahl}, E., {Rappaport}, S., \& {Podsiadlowski}, P. 2003, \apj, 597, 1036,
  \dodoi{10.1086/378632}

\bibitem[{{Skowron} {et~al.}(2019){Skowron}, {Skowron}, {Mr{\'o}z}, {Udalski},
  {Pietrukowicz}, {Soszy{\'n}ski}, {Szyma{\'n}ski}, {Poleski}, {Koz{\l}owski},
  {Ulaczyk}, {Rybicki}, \& {Iwanek}}]{Skowron19}
{Skowron}, D.~M., {Skowron}, J., {Mr{\'o}z}, P., {et~al.} 2019, Science, 365,
  478, \dodoi{10.1126/science.aau3181}

\bibitem[{{Tetarenko} {et~al.}(2016){Tetarenko}, {Sivakoff}, {Heinke}, \&
  {Gladstone}}]{Tetarenko16}
{Tetarenko}, B.~E., {Sivakoff}, G.~R., {Heinke}, C.~O., \& {Gladstone}, J.~C.
  2016, \apjs, 222, 15, \dodoi{10.3847/0067-0049/222/2/15}

\bibitem[{{Veena} {et~al.}(2019){Veena}, {Vig}, {Sebastian}, {Lal}, {Tej}, \&
  {Ghosh}}]{Veena19}
{Veena}, V.~S., {Vig}, S., {Sebastian}, B., {et~al.} 2019, \mnras, 482, 4630,
  \dodoi{10.1093/mnras/sty3032}

\bibitem[{{Whiteoak} \& {Green}(1996)}]{Whiteoak96}
{Whiteoak}, J.~B.~Z., \& {Green}, A.~J. 1996, \aaps, 118, 329

\bibitem[{{Wilms} {et~al.}(2000){Wilms}, {Allen}, \& {McCray}}]{Wilms01}
{Wilms}, J., {Allen}, A., \& {McCray}, R. 2000, \apj, 542, 914,
  \dodoi{10.1086/317016}

\bibitem[{{Xiang} {et~al.}(2018){Xiang}, {Shi}, {Liu}, {Yuan}, {Chen}, {Huang},
  {Wang}, {Wu}, {Tian}, {Huo}, {Zhang}, \& {Zhang}}]{Xiang18}
{Xiang}, M., {Shi}, J., {Liu}, X., {et~al.} 2018, \apjs, 237, 33,
  \dodoi{10.3847/1538-4365/aad237}

\bibitem[{{Xing} \& {Li}(2021)}]{Xing21}
{Xing}, Z.-P., \& {Li}, X.-D. 2021, \apj, 920, 67,
  \dodoi{10.3847/1538-4357/ac16e1}

\end{thebibliography}
\bibliographystyle{aasjournal}



\end{document}